# Realization of quantum spin Hall insulator superlattice with emergent multigap-like helical edge states

Hui Guo[1,2,*], Xianghe Han[1,2,*], Fang Qin[3,*], Xiaoshuai Fu[1,2,*], Hao Peng[1,2], Hongqin Xiao[1,2], Hengxin Tan[4], Chen Liu[5], Zihao Huang[1,2], Jiayi Wang[1,2], Qian Fang[1,2], Rui Chen[6], Haitao Yang[1,2], Wang Yao[7], Li Huang[1,2,†], Hai-Zhou Lu[8,†], Hui Chen[1,2,†], and Hong-Jun Gao[1,2,†]

[1] Beijing National Center for Condensed Matter Physics and Institute of Physics, Chinese Academy of Sciences, Beijing 100190, PR China

[2] School of Physical Sciences, University of Chinese Academy of Sciences, Beijing 100190, PR China

[3] School of Science, Jiangsu University of Science and Technology, Zhenjiang, Jiangsu 212100, PR China

[4] School of Physics and Astronomy, Shanghai Jiao Tong University, Shanghai 200240, PR China

[5] Institute of High Energy Physics, Chinese Academy of Sciences, Beijing 100049, PR China

[6] Department of Physics, Hubei University, Wuhan, Hubei 430062, PR China

[7] New Cornerstone Science Laboratory, Department of Physics, The University of Hong Kong, Hong Kong, PR China

[8] Shenzhen Institute for Quantum Science and Engineering and Department of Physics, Southern University of Science and Technology (SUSTech), Shenzhen 518055, PR China

[*]These authors contributed equally to this work

[†]Correspondence to: hjgao@iphy.ac.cn, hchenn04@iphy.ac.cn, luhz@sustech.edu.cn, lhuang@iphy.ac.cn

**The functional quantum spin Hall insulators (QSHI), protected by time-reversal symmetry against single-particle backscattering, hold great promise for dissipationless quantum electronics. Realization of QSHI with gapped helical edge states, which would enable deterministic on/off switching of the edge-channel conductance, is a key requirement for programmable topological circuits. Here, we report realization of superlattice-modulated QSHI $HfTe_5$ hosting emergent multigap-like helical edge states. Using scanning tunneling microscopy and spectroscopy, we identify a reconstruction-induced periodic superlattice modulation in epitaxial monolayer $HfTe_5$ and directly observe multiple gap-like features in the edge channel, accompanied by a series of sharp peaks in the density of states. Combined with theoretical modelling, we attribute the observed edge gap to the finite-width coupling between the two edges significantly enhanced by the superlattice modulation, whereas the sharp peaks are the manifestations of mini-gaps opening at the reduced Brillouin zone boundaries by the periodic modulation of spin-orbit coupling. Notably, these sharp peaks exhibit clear Zeeman splitting under magnetic fields, consistent with the helical nature of the topological edge states. Our results establish a viable route to engineering gapped helical edge states in QSHI and provide a promising platform for topological devices with desired on/off switchability.**

## Introduction

Helical edge states in quantum spin Hall insulators (QSHI) provide one-dimensional conducting channels with spin-momentum locking, enabling dissipationless charge and spin transport protected by time-reversal symmetry[1–6]. These boundary states not only underpin quantized transport phenomena[2,6–9], but also offer a versatile platform for realizing low-dimensional quantum phases [10-12] and topological functionalities[13-16]. Over the past decade, extensive theoretical studies have been devoted to identifying new QSHI materials[17-20], while experimental observations of helical edge states have been reported in only a limited number of two-dimensional systems, including bismuthene[21], stanene[22], antimonene[23], monolayer $ZrTe_5$[24], ultrathin $Na_3Bi$[25], and monolayer $WTe_2/WSe_2$[7,26,27]. For practical applications, however, a central challenge is the controlled manipulation of these edge channels. In particular, introducing an energy gap into helical edge states would enable deterministic on/off switching of the protected edge-state conductance, providing an essential functionality for topological electronic devices[28-31].

To date, considerable efforts have been devoted to opening gaps in helical edge states. Existing approaches typically rely on strong external perturbations, such as electric or magnetic fields[25,30–36], imperfections or geometric confinement[37,38], or proximity coupling[13,39], which suppress edge conduction by driving topological phase transitions, breaking time-reversal symmetry, or introducing hybridization with external states. However, these strategies either alter the bulk topology or depend on extrinsic conditions, thereby limiting their robustness and controllability. A more desirable and fundamentally distinct approach would be to achieve gap opening while preserving time-reversal symmetry and the bulk topological invariant, thereby enabling precise control of edge-state properties within a given topological phase. Such a capability would allow direct engineering of the edge-state Hamiltonian and open access to a new regime of tunable topological boundary physics. However, realizing this goal remains an outstanding open question.

Here, we report the experimental realization of a QSHI hosting multiple gap-like helical edge states driven by periodic superlattice modulation. By utilizing van-der-Waals epitaxy, monolayer QSHI $HfTe_5$ is synthesized on bilayer graphene/SiC substrate, which exhibits a reconstruction-induced $2\times1$ superlattice modulation. Using scanning tunneling microscopy/spectroscopy (STM/STS), we directly observe robust topological edge states with multiple spectral gap-like features, accompanied by a series of sharp peaks in the differential conductance spectra. Upon applying magnetic fields, these peaks exhibit Zeeman splitting consistent with the helical nature of the edge channels. By combining theoretical modelling, we attribute the edge gap to finite-width interedge coupling, with the superlattice modulation substantially enhancing the gap, while the additional sharp peaks are associated with mini-gaps opened at the reduced Brillouin-zone boundaries by the superlattice modulation of spin-orbit coupling (SOC). Our findings establish a valuable route for engineering energy gaps in topological edge states without destroying the bulk topology, enabling controlled manipulation of edge-state transport.

## Results and discussion

We start with a schematic illustration of gap opening in the topological edge states of a periodic-potential-modulated QSHI (Figure 1a). A pristine QSHI hosts time-reversal-symmetry-protected gapless edge states within the insulating bulk gap, characterized by linear dispersion and a nearly constant density of states (DOS) at finite temperature. When a periodic superlattice potential is introduced, the helical edge states are folded into the reduced Brillouin zone, where band hybridization induced by periodic modulated spin-orbit coupling opens energy gaps. As a result, the originally gapless topological edge states become gapped, accompanied by a strongly modulated DOS at finite temperature (Figure 1a).

We realize such gapped topological edge states in monolayer $HfTe_5$, which has been theoretically predicted to host quantum spin Hall phase[20]. $HfTe_5$ is closely related to $ZrTe_5$, and topological edge states have been experimentally observed at the surface step edges in their bulk forms[40-42]. However, the monolayer form of $HfTe_5$ has not been experimentally realized to date. We successfully synthesize monolayer $HfTe_5$ by van-der-Waals epitaxy on bilayer graphene/SiC(0001) substrates (see Supplemental Materials). The pristine $HfTe_5$ crystallizes in an orthorhombic structure with in-plane lattice constants of $a$=4 Å and $c$=13.8 Å (Figure 1b). The basic building units of trigonal prismatic $HfTe_3$ chains align along the $a$-axis with alternating up- and down-orientations, forming a quasi-one-dimensional striped lattice framework in the $a$-$c$ plane. X-ray photoelectron spectroscopy (XPS) exhibits sharp peaks from Hf *4f* and Te *3d* core levels, with estimated atomic ratio near 1:5, indicating the formation of $HfTe_5$ (Figure 1c). Low-energy electron diffraction (LEED) pattern of the epitaxial $HfTe_5$ shows an additional set of diffraction spots, indicating the formation of a 2×1 superlattice (Figure 1d, S1). Large-scale STM topography of the sample further reveals the formation of monolayer $HfTe_5$ with stripe-like structure (Figure 1e, S2).

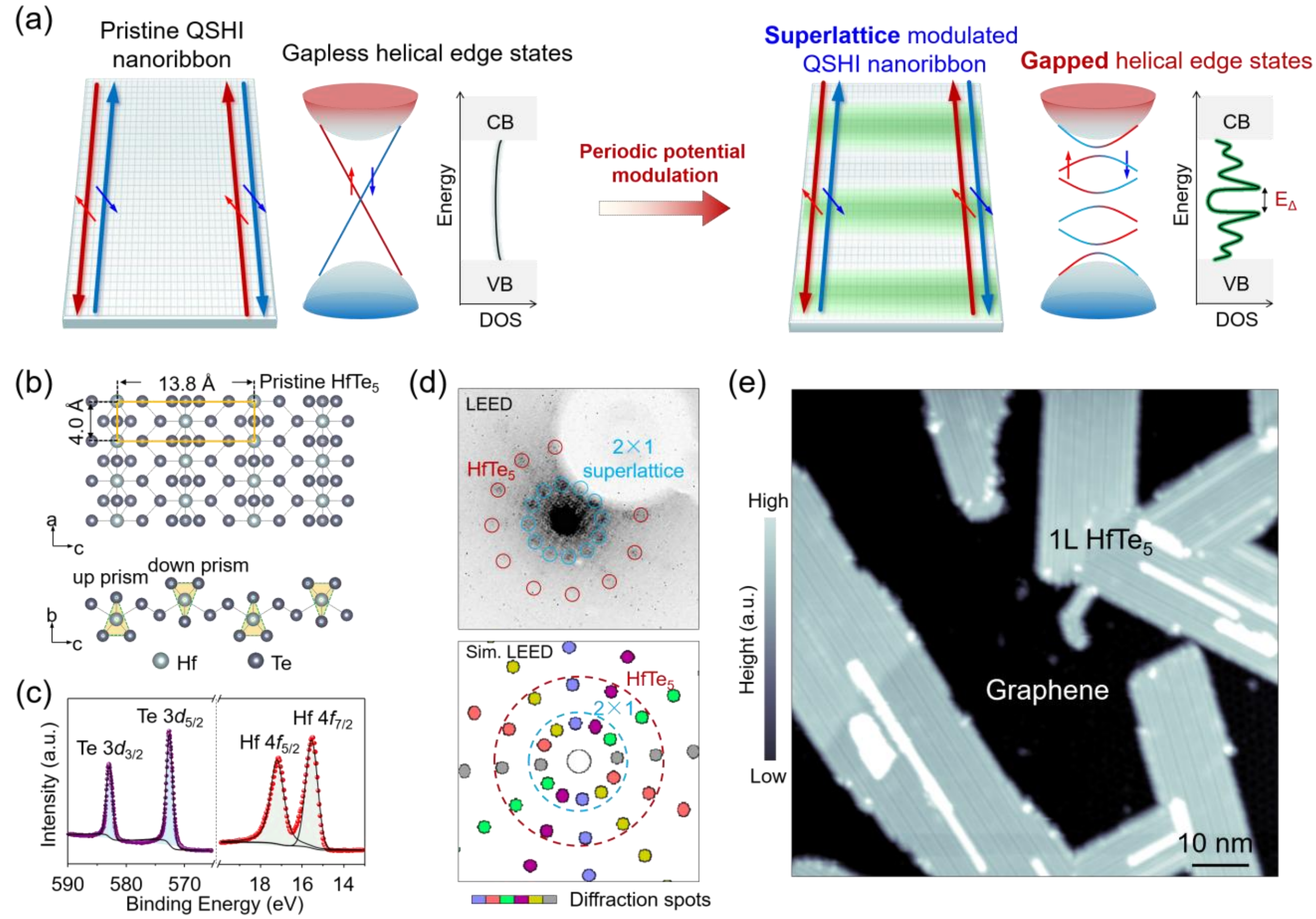

**Figure 1. Schematic of gapped helical edge states and synthesis of monolayer $HfTe_5$.** (a) Left: schematic of a pristine QSHI, showing counterpropagating gapless helical edge states with spin-up (red) and spin-down (blue) dissipationless channels in real space, and corresponding bands in momentum space. Right: schematic of the QSHI with a periodic superlattice potential modulation (green stripe), showing counterpropagating gapped topological edge modes. (b) Top and side views of the pristine $HfTe_5$ crystal structure, showing quasi-one-dimensional arrangement of alternating up- and down-oriented $HfTe_3$ prismatic chains, with lattice constants of $a$=4.0 Å along the chain and $c$=13.8 Å cross the chain. (c) X-ray photoelectron spectroscopy, where the Hf *4f* and Te *3d* core-level spectra are fitted after Shirley background subtraction, yielding an estimated Hf:Te atomic ratio close to 1:5, consistent with the stoichiometric composition of $HfTe_5$. (d) LEED pattern (energy at 23 eV), showing diffraction spots from $HfTe_5$ (red circles) and 2×1 superlattice (blue circles). Lower panel presents the simulated LEED pattern considering 2×1 superlattice and six rotational domains (different colors), showing the 2×1 superlattice, which is in a good agreement with the experimental data in the upper panel. The spots from $HfTe_5$ and 2×1 superlattice are marked by the dashed red and bule circles, respectively. (e) Large-scale STM topography ($V_s$=-2.0 V, $I_t$=0.03 nA), showing the monolayer $HfTe_5$ on graphene substrate.

We then characterize the superlattice structure of the as-grown monolayer $HfTe_5$. Zoom-in STM image of the $HfTe_5$ exhibits four distinct stripes within each unit cell, with a periodicity of 27.3 Å along $c$ direction, twice that of the pristine lattice (Figure 2a). These characters are distinct from the pristine $HfTe_5$ surface that displays two stripes per unit cell associated with two $HfTe_3$ prisms[43-46], suggesting the presence of four $HfTe_3$ prisms in the epitaxial monolayer $HfTe_5$. Atomically-resolved STM and non-contact atomic force microscopy (nc-AFM) images further reveal the atomic structure of the monolayer $HfTe_5$ with a lattice periodicity of ~4.0 Å along *a*-axis, identical to the pristine phase, demonstrating the formation a 2×1 superlattice (Figure 2b, c, S3, S4). Such 2×1 structural reconstruction is further confirmed by the appearance of additional superlattice diffraction spots in the fast Fourier transform (FFT) pattern of the large-area atomically-resolved STM images (Figure S5). Furthermore, theoretically optimized 2×1 $HfTe_5$ comprises two up- and two down-oriented $HfTe_3$ prisms per unit cell, which shows good agreement with the experimental results (Figure 2c). Such a superlattice modulation arises from the reconstruction of the monolayer $HfTe_5$, resembling the results in monolayer $ZrTe_5$[24].

The microscopic origin of the observed 2×1 superlattice is unlikely to be a moiré potential induced by the bilayer graphene substrate. In particular, monolayer $HfTe_5$ domains with different crystallographic orientations relative to the graphene substrate exhibit the same 2×1 periodicity, whereas a moiré superlattice would be expected to depend sensitively on the relative lattice orientation. Moreover, the same superlattice is reproducibly observed in multiple growth batches. Considering the weak van der Waals interaction, negligible lattice-matching constraint, and minimal charge transfer between graphene and monolayer $HfTe_5$, we therefore conclude that the observed superlattice arises from an intrinsic structural reconstruction of monolayer $HfTe_5$.

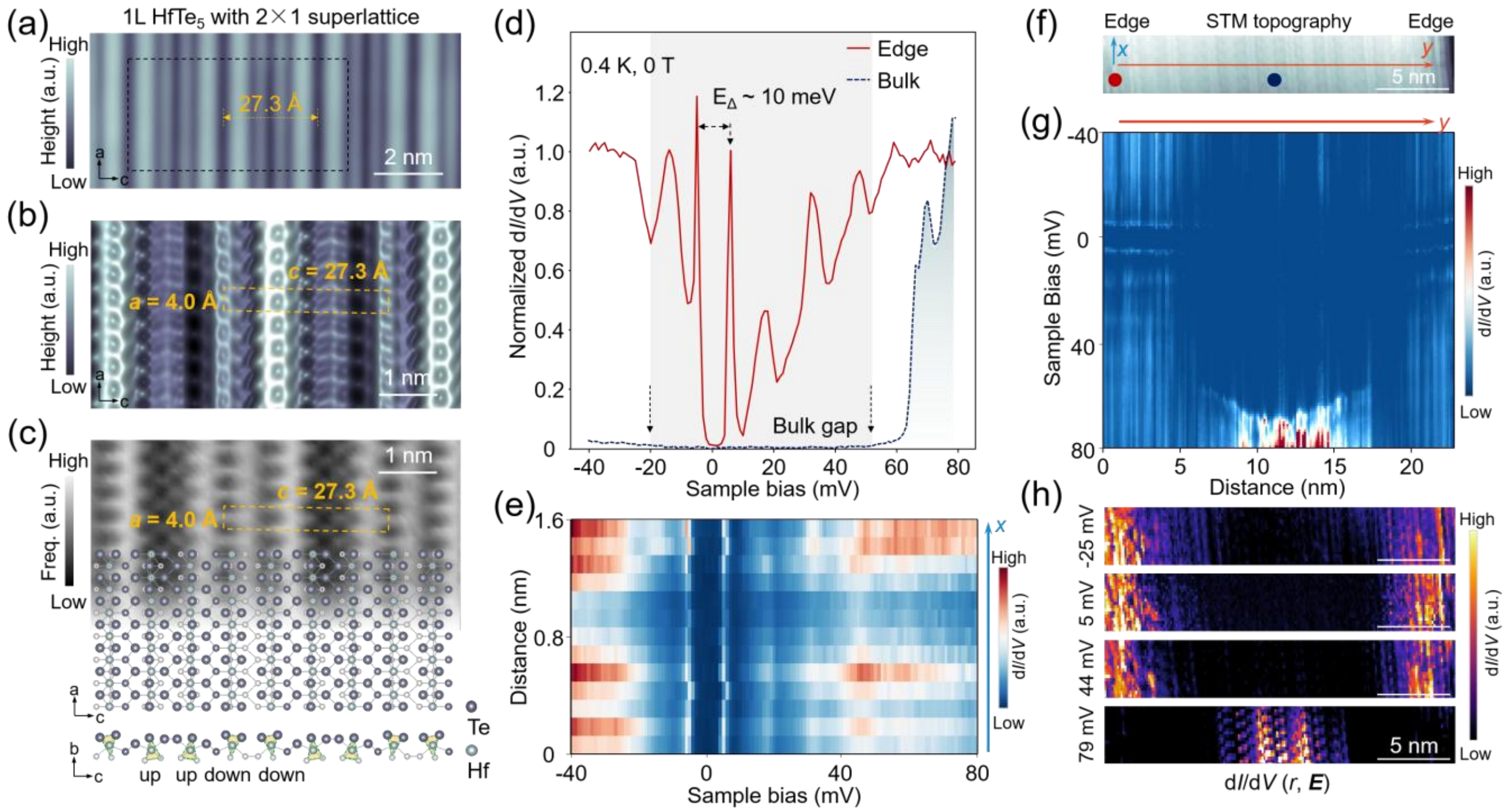

**Figure 2. Observation of multigap-like edge states of superlattice-modulated monolayer $HfTe_5$.** (a) High-resolution STM topography of the monolayer $HfTe_5$, showing quasi-one-dimensional stripe structure with periodicity of ~27.3 Å across the stripe. (b) Atomically-resolved STM image acquired at the black box in (a) ($V_s$=300 mV, $I_t$=0.2 nA), showing lattice constants of ~4.0 Å and ~27.3 Å along *a*- and *c*-crystallographic directions, respectively, indicating a formation of a 2×1 superlattice in the $HfTe_5$ monolayer. The orange dashed rectangle labels the superlattice unit cell. (c) Atomically-resolved nc-AFM image with overlaid the DFT optimized 2×1 $HfTe_5$ structural model, showing a good agreement between the nc-AFM data and DFT calculations results. The crystal structure of 2×1 $HfTe_5$ is composed of two up- and two down-oriented $HfTe_3$ prismatic chains within each unit cell. (d) Typical d$I$/d$V$ spectra acquired at the edge (solid red line) and in the bulk (dashed blue line) of the 2×1 $HfTe_5$ monolayer, showing the emergence of edge states within the bulk energy gap. The edge spectrum exhibits a series of discrete sharp peaks separated by gap-like features, characteristic of a gapped topological edge state. For clarity, the spectra were normalized to the differential conductance outside the bulk gap region (near ~78 mV), which rescales the conductance magnitude and does not affect the spectral features. (e) Color map of the d$I$/d$V$ linecut along the blue arrow in (f), showing uniform energy spacing between adjacent edge-state peaks along the boundary. (f) Zoom-in STM topography of a 2×1 $HfTe_5$ monolayer with width of ~23 nm ($V_s$=0.2 V, $I_t$=50 pA). (g) Color map of the d$I$/d$V$ linecut along the red arrow in (f), showing the bulk energy gap and two similar edge states at double boundaries. (h) Series of d$I$/d$V$ maps of (f) taken at distinct energies, showing the spatial localization of the edge states. For all d$I$/d$V$ data in the figure, tunneling parameter: $V_s$=0.2 V, $I_t$=1.2 nA, $V_{mod}$=0.8 mV.

We further study the topological edge states of the 2×1 superlattice-modulated monolayer $HfTe_5$ ribbon with a width of ~23 nm. At the $HfTe_5$ terrace, the d$I$/d$V$ spectrum exhibits a bulk insulating state (Figure 2d). The bulk gap was determined by linear extrapolation of the leading edges of the valence and conduction bands in the logarithmic-scale STS spectra (Figure S6). The gap values exhibit sample- and position-dependent variation, with representative spectra yielding gap values of approximately 70-110 meV. Spectra acquired at different positions of the same nanoribbon can exhibit different local gap values, indicating that the observed variation cannot be uniquely attributed to ribbon width. This variation reflects the spatially nonuniform electronic structure of finite-width $HfTe_5$ nanoribbons, including the pronounced bending of the bulk-derived bands near the boundaries. Despite this spatial variation, the interior regions consistently exhibit a well-defined bulk insulating gap. These gap values are comparable to the theoretical prediction for monolayer $HfTe_5$[20].

At the ribbon edge, the d$I$/d$V$ spectra shows clear conduction states spanning the entire bulk gap, which is attributed to the topological edge states of the monolayer 2×1 $HfTe_5$ (Figure 2d, S7). Remarkably, a clear energy gap appears at zero bias, with a flat near-zero-conductance region flanked by sharp peaks. In addition, the differential conductance reveals a series of discrete sharp peaks separated by multiple gap-like features, consistent with a multi-gap miniband structure. Both features are distinct from the conventional nearly-flat profiles typically observed for the gapless edge modes[24-26,40-42,47,48], revealing a strongly modulated DOS arising from the superlattice potentials in the 2×1 $HfTe_5$. Furthermore, spatially-resolved d$I$/d$V$ spectra acquired along the edge unambiguously reveal that these sharp resonant peaks are non-dispersive in energies (Figure 2e,f). The d$I$/d$V$ linecut across ribbon's two edges (Figure 2g, S8), together with the d$I$/d$V$ maps recorded at the sharp-peak energies (Figure 2h), consistently show the spatial localization of the multigap-like featured edge states. The spatial decay of several representative edge-state features was analyzed using fixed-bias conductance line profiles measured from the ribbon edge toward the interior (Figure S9). The profiles were fitted to an exponential decay toward the bulk conductance level, yielding decay lengths of approximately 0.6-0.8 nm. These short decay lengths support the spatial localization of the edge-state spectral weight, consistent with previously reported helical edge states[24,41,48]. To verify the reproducibility of the edge state, we further measured both edges of multiple monolayer $HfTe_5$ nanoribbons (Figure S10). All measured edges exhibit nearly similar edge-state spectra with multiple gap-like features, confirming that the observed edge state is robust and reproducible.

It is worth noting that the experimentally observed edge-state features should be understood as spectroscopic gap-like features manifested in the local density of states measured by STS, rather than ideal hard gaps with completely vanishing conductance. Although the differential conductance within the main gap-like feature does not vanish completely, the residual conductance is already comparable to the experimental low-conductance noise level, as confirmed by the logarithmic-scale spectra (Figure S7). Such finite residual spectral weight is expected for an energy scale of only several meV owing to finite quasiparticle lifetime[49], thermal broadening, and experimental broadening[50]. Likewise, the higher-energy minigap-like features appear as reproducible suppressions of the local density of states, likely because the broadened edge-state resonances are superimposed on a finite background conductance.

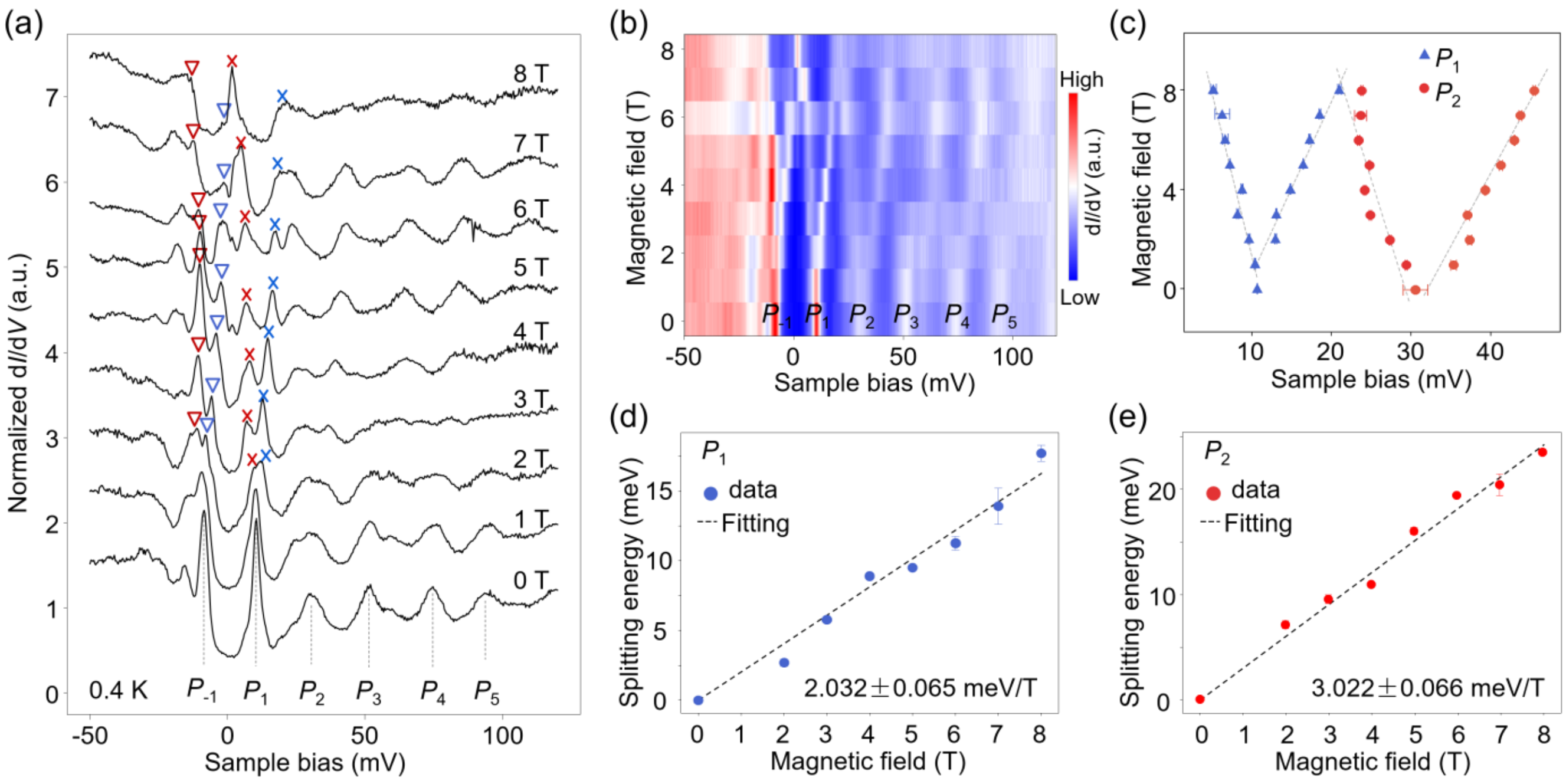

**Figure 3. Evolution of the multigap-like edge states with magnetic field.** (a, b) The d$I$/d$V$ spectra and corresponding color map of the multigap-like edge states under different out-of-plane magnetic fields, showing magnetic-field-induced Zeeman splitting of the edge states. The spectra are vertically offset for clarity The first peak below Fermi level is labeled as $P_{-1}$ and peaks above Fermi level are labeled as $P_1$ to $P_5$. The red and blue triangles mark the two Zeeman-split branches of $P_{-1}$, and the red and blue crosses mark the two Zeeman-split branches of $P_1$. For all d$I$/d$V$ data in the figure, tunneling parameter: $V_s$=0.2 V, $I_t$=1.2 nA, $V_{mod}$ =0.8 mV. (c) Dependence of the energy positions of peaks $P_1$ and $P_2$ on magnetic field. The energy positions for each peak are extracted from Gaussian fitting of the d$I$/d$V$ spectrum at different fields. (d, e) Evolution of the peak-splitting energy $\delta_E$ for $P_1$ and $P_2$ as a function of magnetic field, extracted from (c). Black dashed lines represent linear fits.

To further elucidate the spin properties of the multigap-like edge states, we perform magnetic-field-dependent STS measurements. Under an applied out-of-plane magnetic field, six characteristic peaks ($P_{-1}$, $P_1$-$P_5$) associated with the edge states exhibit clear energy splitting (Figure 3a,b). As the magnetic field $B_z$ increases, each edge-state peak $P_i$ splits into two branches with an energy separation $\delta_i$, forming $P_i \pm \delta_i$ ($B_z$) (Figure 3b). This systematic splitting across multiple peaks indicates a field-driven lifting of spin degeneracy within the reconstructed edge-state spectrum. The observed behavior closely resembles Zeeman splitting, providing direct evidence that the multigap-like edge states retain the spin degree of freedom inherited from the helical nature of the pristine quantum spin Hall edge channels[8,9].

To quantify this behavior, we analyze the field dependence of representative peaks $P_1$ and $P_2$. Their energy positions, extracted by Gaussian fitting of each peak at different fields, exhibit a nearly linear dependence on magnetic field (Figure 3c), consistent with Zeeman-type energy shifts. Furthermore, the two split branches display slightly different slopes, with the higher-energy branch exhibiting a larger slope than the lower-energy one. We further extract the peak-splitting energy ($\delta_E$) for $P_1$ and $P_2$, and plot it as a function of magnetic field (Figure 3d,e), both showing a linear dependence, with small differences in slope indicating distinct effective g-factors for the split states. The slight variation of the extracted effective g-factors likely originates from the fact that different reconstructed edge sub-bands possess different orbital characters and therefore couple differently to the external magnetic field through the spin–orbit interaction. Nevertheless, all peaks exhibit a linear Zeeman splitting with magnetic field, indicating a common spinful origin of the reconstructed helical edge states.

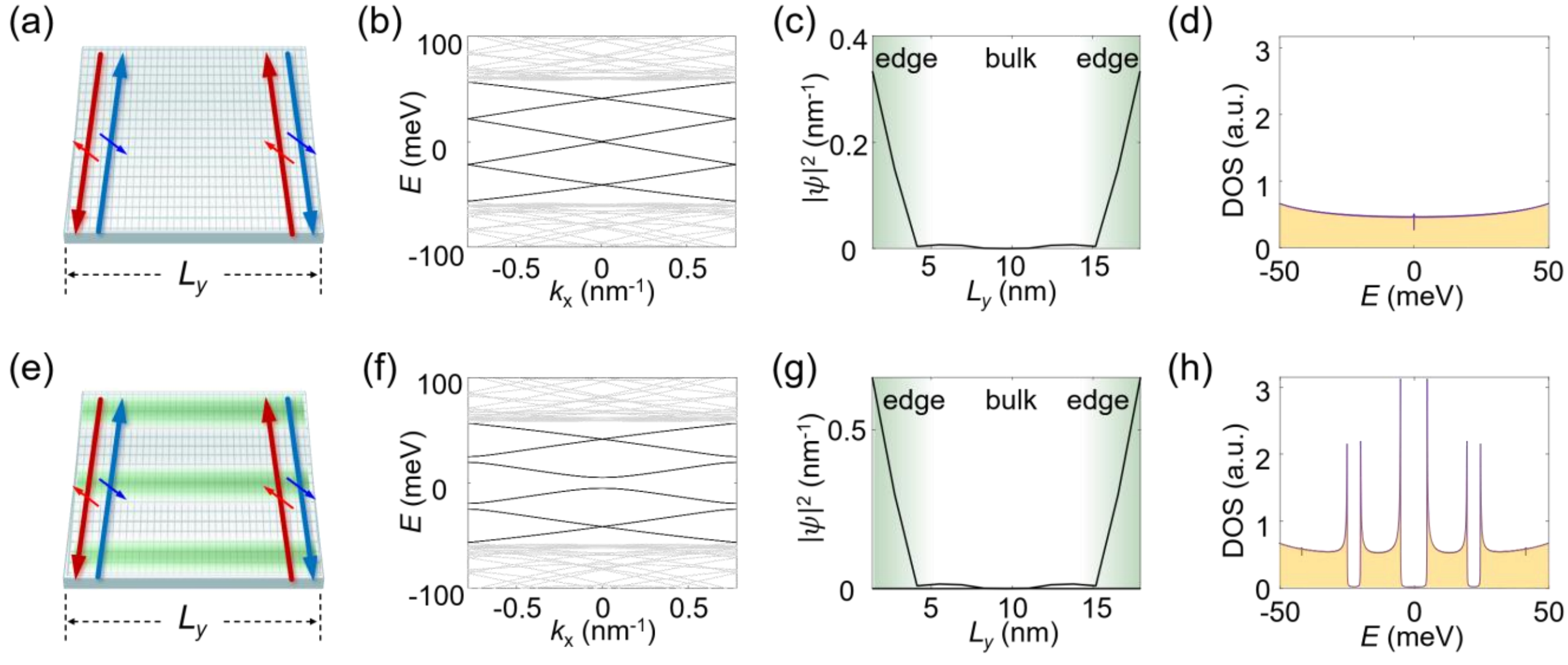

**Figure 4. Periodic-potential-driven mechanism of the multigap-like edge states.** (a)-(d) Schematic of a QSHI absence of periodic potential modulation ($V_x$=$V_y$=0 meV) (a), and the corresponding energy level spectra (b), probability distributions of edge states at $k_x$=0 (c), density of states (d), showing nearly constant density of state of the topological edge states within the bulk gap (100 meV). (e)-(h) Schematic of the QSHI with a periodic potential modulation ($V_x$=$V_y$=5 meV) (e), and the corresponding energy level spectra (f), probability distributions of edge states at $k_x$=0 (g), density of states (h), showing the emergence of multi-gap minibands in the energy level spectrum and multiple gaps with a series of sharp peaks in the density of state. For the calculation, the width of the QSHI is $L_y$=17.94 nm (along y direction), and the lattice constants along $x$ and $y$ directions are $a_x$=0.4 nm, $a_y$=1.38 nm, respectively. Other parameters: $M$=70 meV, $\lambda_x= \lambda_y$=70 meV, $t_x= t_y$=70 meV, $\varphi_x$=0, $N_x$=50, and $\bar{\lambda}_x=n_x a_x$, $n_x$=10.

Finally, we investigate the possible origin of the experimentally observed multigap-like helical edge-state spectrum in a periodicity-modulated QSHI. We first calculated the $Z_2$ topological invariant of the reconstructed monolayer $HfTe_5$. The evolution of the Wannier charge centers yields $Z_2$=1, demonstrating that the reconstructed system remains a quantum spin Hall insulator without undergoing a topological phase transition (Figure S11). Correspondingly, the calculated semi-infinite edge band structure retains an odd topological connectivity across the bulk gap, including a protected Kramers crossing at Γ point. The 2×1 reconstruction may fold the edge bands and gap additional crossings pairwise, but it cannot fully gap an isolated edge while time-reversal symmetry is preserved.

However, the experimentally investigated system consists of finite-width nanoribbons rather than an ideal semi-infinite edge. In a sufficiently narrow finite ribbon, hybridization between counterpropagating edge modes localized at opposite boundaries can produce a central finite-size gap without changing the underlying bulk topological phase. This contribution is expected to decrease rapidly with increasing ribbon width because the wave-function overlap between the two edges becomes exponentially smaller. To examine the possible contribution of finite-width effects, we compared edge spectra acquired from nanoribbons with substantially different widths. As shown in Figure S12, the characteristic central suppression decreases from approximately 19 meV in a 6.7-nm-wide ribbon to approximately 10 meV in a 23-nm-wide ribbon, indicating a considerable width-dependent contribution from finite-width effect. However, given that the edge-state LDOS decays within approximately 1 nm from the boundary, the direct overlap between the two opposing edge states is expected to be strongly suppressed at a ribbon width of 23 nm. Therefore, the residual suppression of approximately 10 meV in the 23-nm-wide ribbon should not be interpreted as a purely interedge-hybridization-induced gap. Moreover, finite-size interedge coupling alone is expected to produce predominantly a single central gap and therefore cannot account for the additional multigap-like features observed in the edge-state spectrum.

To gain deep insight into the underlying mechanism, we construct an effective edge Hamiltonian incorporating the periodic modulation of the effective spin–orbit coupling introduced by the 2×1 structural reconstruction. In the absence of the periodic modulation, finite-width edge hybridization produces only a small gap at the Brillouin-zone center, while the edge-state DOS within the bulk gap remains nearly featureless (Figure 4a-4d). In contrast, introducing a periodic modulation of the effective spin-orbit coupling reconstructs the edge-state dispersion: the central gap is substantially enhanced, accompanied by additional mini-gaps at the boundaries of the reduced Brillouin zone (Figure 4e-4h). Moreover, varying the periodic SOC modulation can change the number, energy positions, and relative spectral weights of the resulting gap-like features (Figure S13), while the overall emergence of a multigap-like spectrum remains robust. The calculated edge-state DOS develops multiple spectral suppressions accompanied by enhanced DOS features, qualitatively resembling the multifeatured edge spectra observed experimentally. These results suggest that finite-width coupling provides the central spectral suppression, while periodic superlattice modulation further reconstructs the edge-state Hamiltonian, enhancing the central gap and generating additional multigap-like features without changing the underlying bulk topological phase.

To further distinguish the role of finite-width effects and periodic SOC modulation, we extended the effective edge-state model to a ribbon geometry covering the range relevant to the experimentally studied nanoribbons and performed systematic calculations as a function of ribbon width. Without periodic SOC modulation, the finite-ribbon calculation produces predominantly a single central suppression (Figure S14). Its characteristic energy scale decreases rapidly with increasing ribbon width, from approximately 19.5 meV at 5.52 nm to approximately 0.3 meV at 17.94 nm, and becomes negligible at 20.70 nm. This behavior is consistent with the exponentially reduced overlap between the edge states localized at opposite boundaries[38]. The finite-size-only spectrum therefore approaches the gapless response of an isolated semi-infinite edge as the ribbon width is increased.

Interestingly, when the periodic SOC modulation is introduced, the edge spectrum is reconstructed and additional finite-energy suppressions emerge (Figure S15). Unlike the pure finite-width induced gap, the central spectral scale remains approximately 9.4 meV for ribbon widths above approximately 11 nm, while an additional reconstructed spectral scale of approximately 5.1 meV remains visible up to 20.70 nm (Figure S16). These characteristic energy scales are comparable to the central and additional suppressions observed experimentally. The controlled comparison between finite-width models with and without periodic SOC modulation reveals the principal distinction between the two mechanisms: conventional interedge coupling produces a rapidly decreasing central-gap feature, whereas periodic SOC modulation substantially enhances the central gap and generates additional width-robust gap-like structures.

It is also important to distinguish the reconstructed edge-state spectrum reported here from gap openings driven by spontaneous electronic ordering. Recently, superlattice formation associated with excitonic or charge-density-wave instabilities has been reported in several quantum spin Hall materials, including monolayer $WTe_2$ and $TaIrTe_4$[51,52]. In these systems, the superlattice originates from a correlated electronic instability and is accompanied by an electronic order parameter. By contrast, the superlattice observed in monolayer $HfTe_5$ is a structural reconstruction. Consistently, Fourier analysis of the edge-state DOS reveals no additional periodic modulation (Figure S17), providing no evidence for an edge-localized electronic order. Moreover, the experimentally observed edge spectrum is well captured by the effective model based on periodic modulation of the spin–orbit coupling without invoking any correlation-driven order parameter. These identify structural superlattice modulation as a fundamentally different mechanism for reconstructing and engineering helical edge states.

Overall, the experimental observations and effective-model calculations provide a consistent mechanism-level picture of the reconstructed edge-state spectrum. The effective model is not intended to quantitatively reproduce the complete STS spectrum of realistic $HfTe_5$ nanoribbons, but rather to provide a physically plausible mechanism for the experimentally observed unconventional edge-state reconstruction. Despite its simplified nature, the model captures several key experimental characteristics and trends, including the emergence of a multigap-like edge spectrum beyond the conventional central finite-size gap, the width dependence of the edge gap, and a characteristic energy scale comparable to that observed experimentally. Moreover, the calculated higher-energy LDOS suppressions become weaker than that of the central one, and incorporating the thermal broadening further reduces the suppression of these higher-energy features (Figure S18), qualitatively consistent with the measured edge spectra. The precise resonance positions, spectral weights, and linewidths remain material-specific and would require a realistic atomistic model of $HfTe_5$ nanoribbons. Nevertheless, the consistency between the calculated spectral characteristics and the experimentally observed trends, together with the ribbon-width dependences, supports the interpretation that the superlattice and finite-width plays an essential role in reconstructing the helical edge-state spectrum.

## Conclusions

We have realized a monolayer superlattice-modulated QSHI nanoribbons and revealed that a periodic potential modulation induces controlled gaps opening in the topological edge states. This modulation gives rise to an intrinsic unconventional helical edge spectrum, featuring multiple discrete gap-like features accompanied by a series of sharp peaks in the density of states, while preserving the bulk topology. The emergence of such edge gaps allows, in principle, on/off control of the one-dimensional topological transport through chemical potential tuning, establishing key switching functionality inaccessible in conventional gapless QSHI systems. More broadly, our results identify periodic potential modulation as an effective mechanism for engineering energy gaps in topological edge states, advancing the controlled manipulation of helical edge transport in low-dissipation topological electronics and computations.

## Methods

**Sample synthesis.** High-quality monolayer $HfTe_5$ nano-ribbons were epitaxially grown on bilayer graphene/SiC(0001) substrates by molecular beam epitaxy (MBE) under ultra-high vacuum conditions (UHV, base pressure ~$5\times10^{-10}$ mbar). The bilayer graphene substrate was prepared repeated flash annealing of the SiC(0001) substrate at 1300 °C after overnight degassing at ~650 °C. During $HfTe_5$ growth, the substrate temperature was maintained within a narrow window of ~175-205 °C. High-purity Hf (99.9%, *ESPI Metals*) and Te (99.99%, *Sigma-Aldrich*) sources were co-evaporated from an electron-beam evaporator and a Knudsen effusion cell, respectively. A Te-rich environment was ensured by maintaining the Te flux approximately one order of magnitude higher than that of Hf.

**Sample characterizations.** After synthesized, the monolayer $HfTe_5$ sample were *in-situ* checked by Low energy electron diffraction (LEED). LEED was employed with a 4-grid detector (*Omicron Spectra LEED*) in an ultrahigh vacuum (UHV) chamber ($1\times10^{-9}$ mbar)) at room temperature, which was equipped in the same chamber of MBE. The XPS measurements were carried out at the photoelectron spectroscopy end-station of the Beijing Synchrotron Radiation Facility 4B9B beamline, using a hemispherical energy analyzer. After growth, the samples were stored in a home-made UHV suitcase and transferred to the XPS chamber for elemental composition analysis. The photon energy of 700 eV was calibrated by referencing the Au*4f* core level of a clean polycrystalline gold foil electrically connected to the sample. The spectra were analyzed by first subtracting a Shirley background, followed by fitting of the core-level peaks using mixed Gaussian–Lorentzian line shapes. The atomic ratio was estimated from the integrated peak areas after applying the corresponding atomic sensitivity factors.

**STM/STS.** After the sample growth, it was then transferred to a *Unisoku*-STM system through a home-made UHV suitcase. STM/STS measurements were performed in an ultrahigh vacuum ($1\times10^{-10}$ mbar) ultra-low temperature condition equipped with 11 T magnetic field. The stable sample temperature can be kept at a base temperature of 0.4 K and 4.2 K, respectively. The electronic temperature is 620 mK at a base temperature of 400 mK. All the scanning parameters (setpoint voltage and current) of the STM topographic images are listed in the captions of the figures. Unless otherwise noted, the differential conductance ($dI/dV$) spectra were acquired by a standard lock-in amplifier at a modulation frequency of 973.1 Hz. Tungsten tip was fabricated via electrochemical etching and calibrated on a clean Au(111) surface prepared by repeated cycles of sputtering with argon ions and annealing at 770 K.

**Non-contact AFM.** The nc-AFM measurements were conducted with a qPlus sensor[53] at 4.5 K in UHV condition. The resonance frequency of the sensor was $f_0$ = 28.46 kHz and the quality factor was q = 50206. The images were taken in constant-height mode by recording the frequency shift ($\Delta_f$) with an oscillating amplitude of 100 pm.

**Theoretical modeling.** To interpret the discrete sharp peaks observed in the topological edge states, we establish an effective theoretical model and calculate the corresponding density of states. Based on Bernevig-Hughes-Zhang (BHZ) model[1] supplemented by a Zeeman energy term $\widehat{\mathcal{H}}_z$ and a periodically modulated spin-orbit coupling potential $\widehat{\mathcal{H}}_{SOC}$ [54], our tight-binding Hamiltonian under periodic boundary conditions along the *x* direction and open boundary conditions along the *y* direction, written in the basis

$$(\hat{C}_{1.1}, \hat{C}_{1.2}, \hat{C}_{1.3}, \cdots, \hat{C}_{1,N_y}, \hat{C}_{2.1}, \hat{C}_{2.2}, \hat{C}_{2.3}, \cdots, \hat{C}_{2,N_y}, \cdots, \hat{C}_{N_x,N_y})^T,$$

is given by

$$\widehat{\mathcal{H}} = \begin{pmatrix} \hat{h}_x + \widehat{U}_x(1) & \widehat{T}_x & 0 & \cdots & \widehat{T}_x^\dagger \\ \widehat{T}_x^\dagger & \hat{h}_x + \widehat{U}_x(2) & \widehat{T}_x & \cdots & 0 \\ 0 & \widehat{T}_x^\dagger & \hat{h}_x + \widehat{U}_x(3) & \ddots & \vdots \\ \vdots & \ddots & \ddots & \ddots & \widehat{T}_x \\ \widehat{T}_x & \cdots & 0 & \widehat{T}_x^\dagger & \hat{h}_x + \widehat{U}_x(N_x) \end{pmatrix}_{N_x \times N_x}, \tag{1}$$

where

$$\hat{h}_x = \begin{pmatrix} \hat{h}_0 + \widehat{V}_y(1) & \hat{t}_y & 0 & \cdots & 0 \\ \hat{t}_y^\dagger & \hat{h}_0 + \widehat{V}_y(2) & \hat{t}_y & \cdots & 0 \\ 0 & \hat{t}_y^\dagger & \hat{h}_0 + \widehat{V}_y(3) & \ddots & \vdots \\ \vdots & \ddots & \ddots & \ddots & \hat{t}_y \\ 0 & \cdots & 0 & \hat{t}_y^\dagger & \hat{h}_0 + \widehat{V}_y(N_y) \end{pmatrix}_{N_y \times N_y}, \tag{2}$$

$$\widehat{V}_y(j_y) = V_y \sigma_x \otimes \tau_x, \tag{3}$$

$$\hat{T}_x = \begin{pmatrix} \hat{t}_x e^{ik_x a_x} & 0 & 0 & \cdots & 0 \\ 0 & \hat{t}_x e^{ik_x a_x} & 0 & \cdots & 0 \\ 0 & 0 & \hat{t}_x e^{ik_x a_x} & \ddots & \vdots \\ \vdots & \ddots & \ddots & \ddots & 0 \\ 0 & \cdots & 0 & 0 & \hat{t}_x e^{ik_x a_x} \end{pmatrix}_{N_y \times N_y} , \tag{4}$$

$$\hat{U}_x(j_x) = \begin{pmatrix} \hat{V}_x(j_x) & 0 & 0 & \cdots & 0 \\ 0 & \hat{V}_x(j_x) & 0 & \cdots & 0 \\ 0 & 0 & \hat{V}_x(j_x) & \ddots & \vdots \\ \vdots & \ddots & \ddots & \ddots & 0 \\ 0 & \cdots & 0 & 0 & \hat{V}_x(j_x) \end{pmatrix}_{N_y \times N_y} , \tag{5}$$

$$\hat{V}_x(j_x) = V_x \cos(q_x j_x a_x + \varphi_x)\, \sigma_x \otimes \tau_x, \tag{6}$$

$$\hat{h}_0 = \begin{pmatrix} M - 2t_x - 2t_y & 0 & 0 & 0 \\ 0 & -M + 2t_x + 2t_y & 0 & 0 \\ 0 & 0 & M - 2t_x - 2t_y & 0 \\ 0 & 0 & 0 & -M + 2t_x + 2t_y \end{pmatrix} + \hat{\mathcal{H}}_z, \tag{7}$$

$$\hat{t}_y = \begin{pmatrix} t_y & -\frac{1}{2}\lambda_y & 0 & 0 \\ \frac{1}{2}\lambda_y & -t_y & 0 & 0 \\ 0 & 0 & t_y & -\frac{1}{2}\lambda_y \\ 0 & 0 & \frac{1}{2}\lambda_y & -t_y \end{pmatrix}, \hat{t}_x = \begin{pmatrix} t_x & -i\frac{1}{2}\lambda_x & 0 & 0 \\ -i\frac{1}{2}\lambda_x & -t_x & 0 & 0 \\ 0 & 0 & t_x & i\frac{1}{2}\lambda_x \\ 0 & 0 & i\frac{1}{2}\lambda_x & -t_x \end{pmatrix} \tag{8}$$

The Zeeman term in Eq.(7) reads

$$\hat{\mathcal{H}}_z = \begin{pmatrix} cg_{\uparrow,1}\mu_B\bar{B} & 0 & 0 & 0 \\ 0 & cg_{\uparrow,2}\mu_B\bar{B} & 0 & 0 \\ 0 & 0 & -cg_{\downarrow,1}\mu_B\bar{B} & 0 \\ 0 & 0 & 0 & -cg_{\downarrow,2}\mu_B\bar{B} \end{pmatrix} \approx cg\mu_B\bar{B}\sigma_z \otimes \tau_0 = M_z \sigma_z \otimes \tau_0, \tag{9}$$

where $\sigma_{x,y,z}$ and $\tau_{x,y,z}$ are Pauli matrices representing the spin and orbital degrees of freedom, respectively, and $\sigma_0(\tau_0)$ denotes the 2×2 identity matrix. Here, $M_z = cg\mu_B\bar{B}$ with $c$=0.974, $\bar{B}$ is the magnetic field intensity, $\mu_B = e\hbar/(2m_e)$ is the Bohr magneton, $e$ is the elementary charge, $\hbar$ is the reduced Planck constant, and $m_e$ is the electron mass. The Landé $g$ factor is taken as $g \approx g_{\uparrow,1} \approx g_{\uparrow,2} \approx g_{\downarrow,1} \approx g_{\downarrow,2}$ for simplicity.

The periodically modulated spin-orbit coupling potential term in Eq.(2) is written as[55]

$$\hat{\mathcal{H}}_{SOC}(j_x, j_y) = \hat{V}_x(j_x) + \hat{V}_y(j_y), \tag{10}$$

where $\hat{V}_y(j_y) = V_y \sigma_x \otimes \tau_x$, $\hat{V}_x(j_x) = V_x \cos(q_x j_x a_x + \varphi_x)\, \sigma_x \otimes \tau_x$, with $V_x$ being the modulation amplitude, $\varphi_x$ the phase, $a_x$ the lattice constant along the *x* direction, $j_x$ indexed the site along the *x* direction, and $q_x = 2\pi/\bar{\lambda}_x$ the wave vector corresponding to the modulation wavelength $\bar{\lambda}_x$.

For supercell approximation, if we choose $n_x$ sites along *x* direction as a supercell, we can have

$$\hat{\mathcal{H}}_{sup} = \begin{pmatrix} \hat{h}_{nx} & 0 & 0 & \cdots & 0 \\ 0 & \hat{h}_{nx} & 0 & \cdots & 0 \\ 0 & 0 & \hat{h}_{nx} & \ddots & \vdots \\ \vdots & \ddots & \ddots & \ddots & 0 \\ 0 & \cdots & 0 & 0 & \hat{h}_{nx} \end{pmatrix}_{(N_x/n_x)\times(N_x/n_x)} + \begin{pmatrix} \hat{U}_x(1) & 0 & 0 & \cdots & 0 \\ 0 & \hat{U}_x(2) & 0 & \cdots & 0 \\ 0 & 0 & \hat{U}_x(3) & \ddots & \vdots \\ \vdots & \ddots & \ddots & \ddots & \hat{T}_x \\ 0 & \cdots & 0 & 0 & \hat{U}_x(N_x) \end{pmatrix}_{N_x\times N_x}, \tag{11}$$

Where

$$\hat{h}_{nx} = \begin{pmatrix} \hat{h}_x & \hat{T}_x & 0 & \cdots & \hat{T}_x^\dagger \\ \hat{T}_x^\dagger & \hat{h}_x & \hat{T}_x & \cdots & 0 \\ 0 & \hat{T}_x^\dagger & \hat{h}_x & \ddots & \vdots \\ \vdots & \ddots & \ddots & \ddots & \hat{T}_x \\ \hat{T}_x & \cdots & 0 & \hat{T}_x^\dagger & \hat{h}_x \end{pmatrix}_{n_x\times n_x}. \tag{12}$$

The density of state is calculated as[56]

$$D(E) = \frac{1}{N_{k_x}} \sum_{k_x,i} f(E - E_{k_x,i}), \tag{13}$$

where *E* is the probe energy, $E_{k_x,i}$ denotes the *i*th eigenvalue for the fixed $k_x$ of the Hamiltonian (1), $N_{k_x}$ is the total number of discrete $k_x$ with $k_x \in [-\frac{\pi}{n_x a_x}, \frac{\pi}{n_x a_x}]$, and $f(E - E_{k_x,i})$ is given by[57]

$$f(E - E_{k_x,i}) = \lim_{k_B T \to 0} \frac{1}{\pi} \frac{k_B T}{(E - E_{k_x,i})^2 + (k_B T)^2}, \tag{14}$$

with the temperature *T* and the Boltzmann constant $k_B$.

**Acknowledgements**

The authors are grateful to Hongming Weng, Ziqiang Wang, Yuanfeng Xu, and Haowei Chen for insightful discussions. This work is supported by grants from the National Natural Science Foundation of China (62488201, 52572188, 92580202, 12525401, 12304195), the National Key Research and Development Projects of China (2022YFA1204100, 2022YFA1403700), the CAS Project for Young Scientists in Basic Research (YSBR-053 and YSBR-003). F.Q. acknowledges support from the Jiangsu Specially Appointed Prof. Program in Jiangsu Province and the Doctoral Research Start-Up Fund of Jiangsu University of Science and Technology.

**Author Contributions**

H.G., H.C. and H.-J. G. designed the experiments. H.G., H.P., H.X., J.Y.W., H.T.Y. and Q.F. fabricated the samples. C.L. did the XPS measurements. X.F. and L.H. performed nc-AFM characterizations. X.H., H.C., and Z.H. performed STM experiments. H.X.T. did the DFT calculations. F.Q., R.C., and H.Z.L. performed the theoretical modelling. All of the authors participated in analyzing the data, plotting figures, and writing the manuscript. H.-J.G. supervised the project.

**Data Availability Statement**

The data that support the findings of this study are available from the corresponding author upon reasonable request.

**Competing Interests:** The authors declare that they have no competing interests.